\relax
\documentclass[letterpaper]{article} 
\usepackage{aaai20}  
\usepackage{times}  
\usepackage{helvet}  
\usepackage{courier}  
\usepackage[hyphens]{url}  
\usepackage{graphicx}  
\usepackage{subfigure}
\usepackage{graphicx}
\usepackage{amsthm}
\usepackage{mathrsfs}
\usepackage{makecell}
\usepackage{amsmath}
\usepackage{amssymb}
\usepackage{bm}
\usepackage{textcomp}
\usepackage{algorithm}
\usepackage[noend]{algpseudocode}
\usepackage{subfigure}
\usepackage{booktabs}
\usepackage{nameref}
\usepackage{mathrsfs}
\usepackage{multirow}
\usepackage[small]{caption}
\usepackage{paralist}
\usepackage{xcolor}

\begin{document}
\title{Addressing Time Bias in Bipartite Graph Ranking for Important Node Identification}
\author{
Hao Liao\textsuperscript{1},
Jiao Wu\textsuperscript{1},
Mingyang Zhou\textsuperscript{1},
Alexandre Vidmer\textsuperscript{1}
\\
\textsuperscript{1}{ College of Computer Science and Software Engineering, Shenzhen University, Shenzhen 518060, China}
}
\maketitle
\begin{abstract}
The goal of the ranking problem in networks is to rank nodes from best to worst, according to a chosen criterion. In this work, we focus on ranking the nodes according to their quality.
The problem of ranking the nodes in bipartite networks is valuable for many real-world applications. For instance, high quality products can be promoted on an online shop or highly reputed restaurants attract more people on venues reviews platforms.
However, many classical ranking algorithms share a common drawback: they tend to rank older movies higher than newer movies, though some newer movies may have a high quality. This time bias originates from the fact that older nodes in a network tend to have more connections than newer ones.
In the study, we develop a ranking method using a rebalance approach to diminish the time bias of the rankings in bipartite graphs.

As a result, it is indicated that the rankings produced after applying the rebalancing of the scores not only reduce time bias, but also reflect the quality of the benchmark items more accurately. The ranking performance improvements range from 20\% to 80\%, the experiments were conducted on three real datasets with ground truth benchmark.

\end{abstract}

\section{Introduction}
In online social networks, we constantly need to select content, for instance which content to read, which videos to watch, or which people to follow~\cite{zheng2018drn}~\cite{pinto2013using}. However, our time is limited, and the amount of content available is so huge that we have no time to go through all of them~\cite{Chen:2017:PKF:3077136.3080776}. Then, we rely on rankings by algorithm or people to guide us to this huge volume of data~\cite{he2018nais}. The results of these rankings influence our decision on which products to buy, which Web pages to visit, whom to choose as a friend, and so on~\cite{wang2018reinforcement}~\cite{szabo2008predicting}~\cite{Chen:2017:ACF:3077136.3080797}. It is then pretty clear that the rankings are very important in our daily life.

Producing the ranking is the first step, but then it is needed to evaluate the quality of the produced ranking.
It is usually a difficult task due to the lack of real \textit{ground truth} data. The traditional assessment
of information quality is based on the discernment of a few selected experts. For instance, Oscars are awarded
to movies by a comity of people working in the movie industry. The awarded movies can then be used to form
a list constituting a kind of \textbf{ground truth} for our evaluation.
Still, a  question remains: are algorithms able to compute the intrinsic quality of movies?

In this work, the \textbf{network/graph} representation is used. A network is composed of nodes and links,
which represent objects and their interactions. A link connects two nodes if they have an interaction.
For instance, in the citation network, nodes are scientific papers and links are citations. Bipartite
graph is a special case in which two types of nodes are present, and the link can only present the
interaction between two different types of nodes. For instance, in Netflix, one type of node is the user
and the other one is a movie. For simplicity, we refer to one type of node as \textbf{user} and the other one as \textbf{item}
in the rest of the paper. The network framework provides a great tool to represent the relationship between users and items.
With online data, it is easy to form a bipartite graph that represents the interaction
between nodes~\cite{cao2011bipartite}.
An important goal in the bipartite graph ranking is to find the high quality nodes based on a specific definition of what is
quality~\cite{cai2017hybrid}. The quality can for instance reflects the likeliness of an item to be popular in
the future~\cite{feng2017computational}.
It can also be used to perform link prediction and recommendation for the users~\cite{liao2018attributed}. Neural networks have recently successfully been applied to the recommendation problem in bipartite network \cite{lian2018towards,wang2018reinforcement}.

The most famous ranking algorithms mainly deal with unipartite graphs. Those algorithms including:
PageRank~\cite{ilprints422}, HITS~\cite{Kleinberg:1999:ASH:324133.324140}, and their variants. The bias towards old nodes in networks has been studied for PageRank in the scientific citation network \cite{mariani2016identification}. Similarly in the movie network, there is a bias toward older items with the classical ranking algorithm. This is not in the interest of the users to have a ranking composed mainly by old movies.

In this paper, we propose a solution to the time-bias problem of ranking nodes in bipartite networks.
We use the structural property of the network and the features of the nodes, including rating, degree, and time of appearance in the network. The main highlights of this paper are the followings:
\begin{itemize}
\item We develop a new method to balance the ranking results in bipartite networks according to their age, and show the impact of this balance process on the ranking.
\item We compare our method with state-of-art ranking methods, and present a complementary evaluation with the external ground truth. These experiments enable future applications to adapt to this important balancing procedure.
\item We compare eight algorithms on three widely-used datasets, illustrating how our method enhances the identification rate of high quality vertices, and also enhance the balance of the ranking results.
\item We study the influence of the parameters on the performance, proving our method does not depend on the fine-tuning of parameters.
\end{itemize}
The remaining paper is organized as follows. We first introduce related works, graph-based ranking methods, and bipartite graph-based ranking methods in \nameref{sec:related}. We describe the network approach and the general problem in \nameref{sec:problem}. Then we describe our approach in \nameref{sec:balance}, and in \nameref{sec:expe} we describe the datasets, the metrics, and the other algorithms used in this work. The results of our method and the comparison with other methods is shown in \nameref{sec:results}.
Finally, we summarize and discuss our main results in \nameref{sec:conclusion}.

\section{Related Work}
\label{sec:related}

Ranking items or people is a widespread practice. The method developed in this work aims to rank the objects in graphs. In this section, we will then review the preceding works which also used the graph approach to rank objects.

One of the most famous ranking algorithms is PageRank~\cite{ilprints422}, which ranks Web pages according to their importance in the network. This algorithm was used to provide search results by Google's search engine. This example illustrates the importance of the ranking algorithms in online networks, as it became one of the most important companies in the world. The simplicity of this algorithm allowed it to be modified and used for various situations and ranking goals. For instance, Reverse PageRank~\cite{fogaras2003start} gives a high score to Web pages that are highly connected in the network. These Web pages are critical in the network and their removal would highly affect the network's connectivity. Topic sensitive PageRank improved the original PageRank algorithm by adding context into the algorithm query~\cite{haveliwala2002topic}.
Hub, authority and relevance scores algorithm (HAR)~\cite{li2012har}, extended the concept of PageRank to graph with multi-relational data.


Another famous algorithm was created in the same period: the HITS algorithm~\cite{Kleinberg:1999:ASH:324133.324140}, which also aimed to rank Web pages according to their importance in the network. Both the methods are iterative and are based on the fact that important Web pages have links pointing to other important Web pages. The main difference is that HITS algorithm consists of two scores: one authority score, and one hub score. The fact that this algorithm uses two different scores makes it a good candidate to be used in bipartite graphs~\cite{deng2009generalized,he2017birank}.

In this work, we are interested in the ranking of nodes in bipartite graphs. As said above, the HITS algorithm was extended to the bipartite graphs. The first extension is co-HITS~\cite{deng2009generalized}. The considered bipartite graph was the graph composed of queries and Web pages. A user submits a query to the engine, and the co-HITS algorithm shows the results according to their computed relevance. More recently, the BiRank algorithm \cite{he2017birank} was developed to be used on any type of bipartite networks, for instance, queries and Web pages, or users and items. This algorithm is iterative as PageRank and HITS, but the normalization is performed differently.


\section{Problem Statement}
\label{sec:problem}

We use the network approach to define our problem. A graph/\textbf{network} $G(V, E)$ is a set of vertices/nodes $V$ connected by edges/links $E$. The nodes $v\in V$ are the objects of the networks. It can for instance represent people, scientific publications, or music albums. The edges $e\in E$ are the connection between the nodes. Their meanings depend on the context: for instance, if the nodes are people, an edge could indicate a friendship relation.
\begin{figure}[!htbp]
\centering
\includegraphics[width=\linewidth]{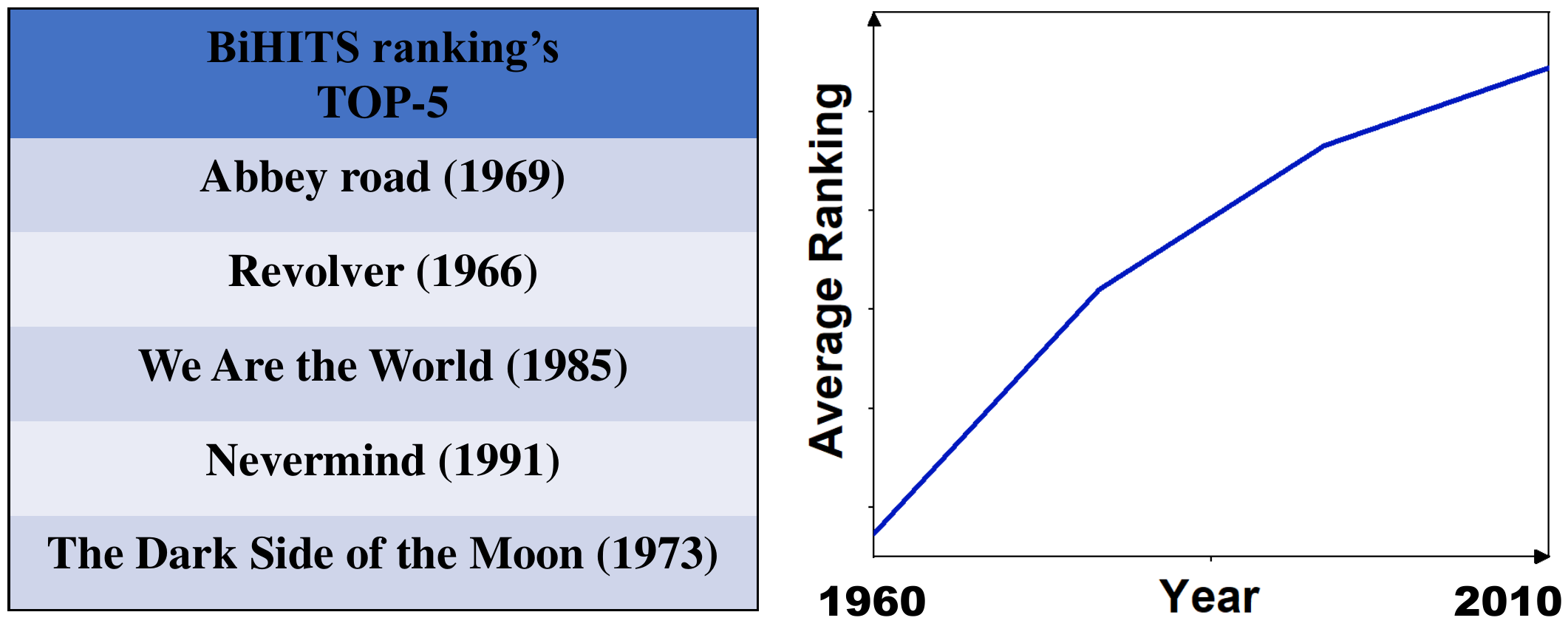}
\caption{Illustration of the time bias obtained with the BiHits ranking algorithm on Amazon dataset (see Section~\nameref{sec:balance}). On the left the five albums with the highest score, and on the right the average ranking position on the y-axis as a function of the release year on the x-axis.}
\label{fig:time_biased}
\end{figure}

In our problem, the graphs are said to be \textbf{bipartite}. A graph is bipartite if it can be divided into two sets of nodes such that each node is only connected to nodes in the other sets. In this work, one set of nodes are \textbf{users}, and the others are \textbf{items} (movies or music albums in this work). A link connects a user and an item if the user has rated the item. In the mathematical formulas and illustrations, we use Latin letters subscripts for users, and Greek letters subscripts for items.
Our rebalance method is motivated by the time basis present in classical ranking algorithms (as shown in Figure \ref{fig:time_biased}). This bias is due to the fact that
in many networks, popular nodes tend to get more attention than less popular nodes, thus favoring older nodes over newer ones \cite{price1976general}.
When there is such a strong bias in the ranking process of an algorithm, it is obvious that it cannot be reliable at estimating the intrinsic quality of nodes. It is then necessary to address this problem by balancing the scores attributed to nodes
For a ranking list to be useful to users, it should not only display well-known items; the more recent items should also have the opportunity to make it to the top of the list.
In order for users to really benefit from the ranking list, we not only need to find good classical well-known projects, but also let excellent works of different periods of time have the opportunity to present to users.
The ranking algorithms considered here only rely on the structure of the network, without the use of metadata or external additional data: given a bipartite network $G(V, E)$, a ranking algorithm produces a score list $R$, where each $R_\alpha$ in the list represents the quality of the item obtained with the algorithm. The quality of the score list is assessed by comparing it with the opinion of experts: a good score list should assign a higher score to the items that have been praised by the experts. The problem we address in this work is the time-base of rankings: usual ranking algorithms give a higher score to older nodes compared to newer ones. Then, the top of the rankings obtained using such algorithms are occupied by old items~\ref{fig:time_biased}.

\section{Proposed Framework}
\label{sec:balance}
We provide a method that assigns high scores with the same frequency to old and new nodes. For online shops, new products are easier to be discovered and bought by customers. In the art world, it makes more recent works more likely to be noticed and appreciated.
\begin{figure*}[!htbp]
\includegraphics[width=.95\linewidth,height=3.5in]{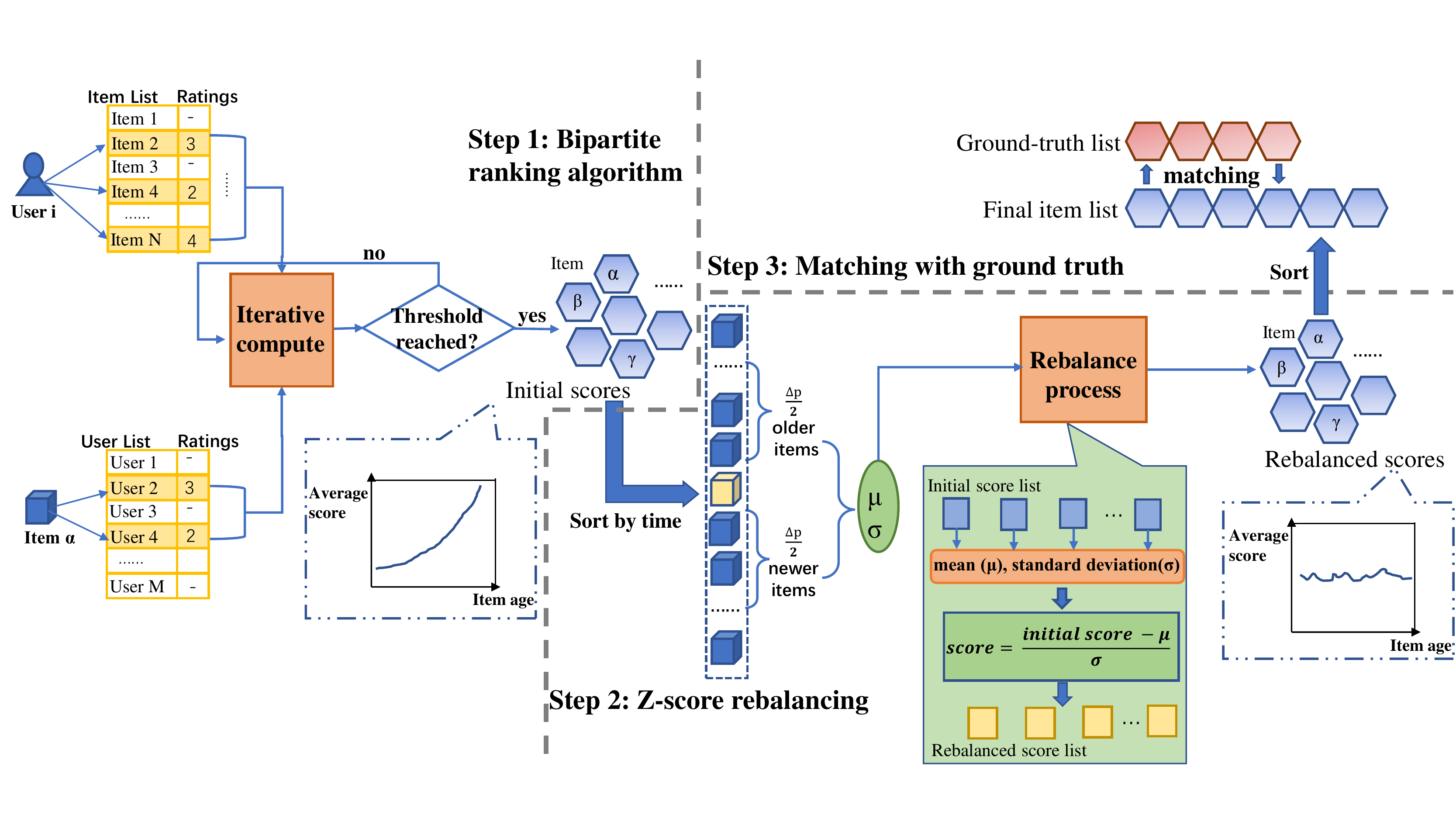}
\caption{Diagram of the method. The inputs of the method are the ratings of the items by the users, and the output is a list of score for each item in the network.   }
\label{fig:illustration}
\end{figure*}

\subsection{Time biased ranking algorithm}
We design our framework so that any ranking algorithm can be used as a baseline method to obtain the scores. Several algorithms were tested in this work, and \textbf{BiRank} stood out as the method of choice~\cite{he2017birank}. The baseline performance of this algorithm were among the best, and its conception allows it to be easily extended with additional data such as query, or prior knowledge on users/items.
The equations read:
\begin{equation}
R_i=\sum_{\alpha}{\frac{w_{i\alpha}}{\sqrt{d_i}\sqrt{d_\alpha}}}F_\alpha
\label{equa:birank_r}
\end{equation}
\begin{equation}
F_\alpha=\sum_{i}{\frac{w_{i\alpha}}{\sqrt{d_\alpha}\sqrt{d_i}}}R_i,
\label{equa:birank_f}
\end{equation}
where $R_i$ and $F_\alpha$ are the score for user $i$ and item $\alpha$, respectively, $d_i$ and $d_\alpha$ their degree, and F$w{_i\alpha}$ denote the weight between user $i$ and item $\alpha$. In this work, we use two different kinds of weights, the ratings and the time decaying weights
   $ w_{ij}=\delta^{a({t}-{t_{ij}})}$,
where $\delta$ is the time decay parameter, $t_{i\alpha}$ is the time at which user $i$ reviewed item $\alpha$, $t$ is the current time, and $a$ is a parameter to control the decay. In this works, we set the value of $\delta$ to $0.85$, $a$ to $1/1 year$. The fine tuning of the decay parameters does not affect the results significantly.
We label them BiRank$_r$ and BiRank$_t$, respectively. The algorithm is iterative, the users and items scores are initialized from uniform random distribution between 0 and 1. The iterations stop when the difference between two consecutive steps is smaller than a threshold: $\| \mathbf{R}_i - \mathbf{R}_{i-1}\| < th$ and $\| \mathbf{F}_i - \mathbf{F}_{i-1}\| < th$.

BiRank method uses the bipartite graph's structure and the user and item's previous information, distributes score to nodes iteratively, and finally converges to a stable and unique ranking result. BiRank smooths the edges' weight by the degree of two linked nodes while using symmetric normalization is one of the key characteristics of the method. However, the scores obtained with this method are time biased.

\subsection{Standardization of the scores}
The crucial step in our method is the use of this standardization to rescale scores.
There are different ways to rescale scores. For instance, one could  rescale the scores with min-max scaling $F^\prime_\alpha = (F_\alpha - \min_\beta(F_\beta) / (\max_\beta(F_\beta)\min_\beta(F_\beta))$. However, this type of normalization is sensitive to outliers, as minimum and maximum are only determined by one small or large value in the sample.
However, in our problem the items are reviewed in different time periods, and the context is different for each time period.
The user's reviews will be time dependant. Obviously: users in 2000 and in 2015 have different opinions and criteria.
As we are interested in the comparison between different sample, we need a method which make samples comparable. The z-score is suitable for this, as it allows for a greater comparability of the data: each standardized set has a mean of zero and a standard deviation of one. Even if the mean and the standard deviation of scores vary over time, this process allows a fair comparison between all items. The z-score is defined as:
\begin{equation}
z={(x-\mu)}/\sigma
\end{equation}
where $\mu$ denotes the average value of the population, and $\sigma$ denotes the standard deviation of the population. When the z-value is negative, it means that $x$ is below the average value; when it is positive, it means that $x$ is above the average value. Moreover, the value of $z$ is equal to the number of standard deviation between the value of $x$ and the mean.

In our case, the scores obtained by the ranking algorithms are standardized over time, similarly to \cite{mariani2016identification}. The process is the following. We first compute all items' scores $\mathbf{F}$ with a ranking algorithm. This base score can be computed from any ranking algorithm. Then, we sort all items according to the time at which they enter the network. In order to compute the score of item $\alpha$,  we select p close items according to the time at which the item was published/released. The rebalance score of item $\alpha$ is then:
\begin{equation}
F^\prime_\alpha(p_\alpha)=\frac{F_\alpha-u(p_\alpha)}{\sigma(p_\alpha)},
\end{equation}
where $u(p)$ is the average score in $p$ items, $\sigma(p)$ the standard deviation and $p_\alpha$ the close set in which item $\alpha$ belongs to.

The rebalance process removes the rewards or penalties on items' scores due to the time of period at which they appear. It is important to note that the selection of the items appearing at the same time period is defined with a number of items.
Several different approaches on how to group items were tested. For instance, grouping the items that enter in the network on the same day, week or month did not lead to satisfactory results.
Finally we decided to group the items with the closest $\Delta p$ items in time. For item $i$, its group is composed by the $\frac{\Delta p}{2}$ items that are older, and the $\frac{\Delta p}{2}$ items that are newer. In the case there is less than $\frac{\Delta p}{2}$ items older than $i$, we choose the $\Delta p$ oldest items, and similarly for new items. The method is illustrated in Figure~\ref{fig:illustration}.

\section{Experimental Setting}
\label{sec:expe}
In this section, we test the balance framework coupled with five algorithms. Three real datasets are used: Amazon Music, MovieLens, and Netflix. Their basic statistics are illustrated in Table~\ref{tab:dataset}. Then we show the performance of the balance framework with different metrics to prove its ability to identify important nodes and to time balance the scores.

\subsection{Datasets}
\begin{table}[!htbp]
\small
\centering
\setlength{\tabcolsep}{1.0mm}
\caption{\textbf{Datasets Basic Statistics, with $k_i$ the average degree of users and $k_\alpha$ the average degree of items. }}
\centering
\begin{tabular}{ c | r | r | r | r | c  }
	\toprule
	\textbf{Name} & \textbf{\makecell{\#Users}} & \textbf{\makecell{\#Items}} & \textbf{\makecell{$<k_i>$}} & \textbf{\makecell{$<k_{\alpha}>$}} & \textbf{Sparsity} \\
	\midrule
	\ Amazon Music&$22959$&$62828$&$31.77$&$11.61$&$5.05\times10^{-4}$  \\
	\ MovieLens&$138395$&$2101$&$65.95$&$4344.26$&$3.14\times10^{-2}$\\
	\ Netflix&$478692$&$5892$&$147.26$&$11963.67$&$2.50\times10^{-2}$\\
	\bottomrule
\end{tabular}	
\label{tab:dataset}
\vspace{-1em}
\end{table}
\subsubsection{Amazon Music}
 Amazon is the biggest online shopping platform in America. The Amazon Music data used in this work were gathered by Stanford University's SNAP group\footnote{http://snap.stanford.edu/}~\cite{leskovec2007dynamics}. We firstly clean the data: delete those unused content from metadata, and extract user ID, product ID, and the timestamps of the ratings. The ratings range from 1 to 5. Some albums come in different versions. In order to solve the problems that could arise from these duplicate items, we combine the reviews of users on these items.
 Finally, to ensure the validity of the data, we only select the users with at least 10 reviews as valid users and products that received at least 10 reviews as valid products.
\subsubsection{MovieLens}
The MovieLens dataset was collected from a well-known movie recommendation website, which was founded by the GroupLens research team of the School of Computer Science and Engineering at the University of Minnesota\footnote{https://grouplens.org/datasets/movielens/} \cite{harper2016movielens}. We only choose the users that have reviewed at least 20 movies as valid users, and choose those movies that received more than 20 reviews as valid items. User's ratings for movies range from 1 to 5. The dataset does not provide the release date for movies. In order to obtain it, we mapped metadata with the MovieLens dataset.
\subsubsection{Netflix}
The company Netflix released data about its users' records and organized a data science competition, whose goal was to predict the users' ratings of movies \footnote{https://netflixprize.com}. Since then, the Netflix dataset has been used for various purposes, such as recommendation, or ranking of movies in our case. The original data contains around 18'000 movies and 500'000 users. Note that for Netflix the release year of movies is available, but not the release date, which is essential in our study. After the mapping process for the release dates, we are left with 5'892 movies.

\subsubsection{Ground truths}
In general ranking problem, it is difficult to define whether a product has higher intrinsic quality because different users will have different views and feelings about it, the quality of the product depends on many intangible subjective factors. We can't simply judge a product's quality according to some good reviews or low ratings. In this paper, we choose specific products that are widely recognized to be of high quality to build the ground truth. Both our ground truth data can be collected on public websites. \newline
\noindent \textbf{Amazon Music} For the Amazon Music dataset, we collect albums that have won the Grammy Music Awards to build the list of high quality items used for the ground truth. The Grammy Awards awards music that has achieved outstanding results over the past year. The voting committee is composed of members of the entertainment industry. For the music that won this award, we consider that they are widely recognized. \newline
\noindent \textbf{MovieLens and Netflix} For the movie datasets, we choose the movies that won the Oscar Award to be the ground truth items. In the movie industry, the Oscar is the most significant award in the world. A film, which was awarded at the Oscars, is a work recognized by the audience around the world as well as filmmakers and is well qualified as a ground truth.
Additionally to the Oscar awards, we selected the top250 ranking from IMDb. This ranking is produced by IMDb's own algorithm based on users' reviews, but it is a very famous top and we use it as an additional ground truth for our algorithms. The main interest of this top is that, unlike the Oscar awards, it is not evenly distributed over the years.

\subsection{Metrics}
\noindent \textbf{Imbalance}
We propose a metric to quantitatively analyze time balance about the ranking result. According to the assumption of~\cite{radicchi2008universality,radicchi2012testing}, with an ideal ranking algorithm, old nodes and new nodes should have equal opportunity to get high ranking. For example, divide all nodes into $S$ groups ${G_1, G_2, G_3,..., G_S}$ according to the time that they enter the network, in which $G_1$ is the earliest group and $G_S$ is the latest group. With a time balanced ranking algorithm, items in every group have the same probability to rank in the Top-$L$ position (such as Top-$1\%$), which means the number of items that rank in Top-$L$ in every group should obey the hypergeometric distribution~\cite{radicchi2012testing}. We define $m$ the number of items, the average number of items that enter Top-$L$ per group is $n^{(0)}=mL/S$. Then the standard deviation of the ideal ranking algorithm is:
\begin{equation}
\sigma_0(L)=\sqrt{\frac{mL}{S}(1-\frac{1}{S})(1-L)\frac{m}{m-1}}
\label{(11-1)}
\end{equation}

\begin{table*}[!htbp]
\centering
\setlength{\tabcolsep}{1.0mm}
\caption{
Performance of the algorithms on the Awards for the three datasets. The Groud truth of Amazon Music is Grammy, and for MovieLens and Netflix are Oscar and IMDb Top 250, respectively. The ranking lists are composed of the top-1\% items for each algorithm. Here Rec. stands for Recall and Pre. stands for Precision.
}
\makebox[\textwidth][c]{
\begin{tabular}{ c | r | r | c | r | r | c | r | r | c | r | r | c | r| r| c}
\toprule
\multirow{2}{*}{\textbf{Method}} & \multicolumn{3}{c|}{\textbf{Amazon Music }} & \multicolumn{3}{c|}{\textbf{MovieLens(Oscar)}} & \multicolumn{3}{c|}{\textbf{Netflix(Oscar)}} &
\multicolumn{3}{c|}{\textbf{MovieLens(IMDb)}} & \multicolumn{3}{c}{\textbf{Netflix(IMDb)}}\\
& \makecell{Rec.} & \makecell{Pre.} & \makecell{AUC} & \makecell{Rec.} & \makecell{Pre.} & \makecell{AUC} & \makecell{Rec.} & \makecell{Pre.} & \makecell{AUC} &
\makecell{Rec.} & \makecell{Pre.} & \makecell{AUC} &
\makecell{Rec.} & \makecell{Pre.} & \makecell{AUC}\\
\midrule
BiHITS &  $.326$ & $.72$ & $.932$ & $.318$ & $.333$ & $.977$ & $.62$& $.271$	& $.738$ & $.219$& $.762$	& $.834$ & $.170$& $.305$	& $.892$ \\
RB-BiHITS & $.478$ & $.105$   & $.928$ & $.455$& $.476$ & $\textbf{.979}$ & $.101$& $.441$	& $.824$ & $\textbf{.247}$& $\textbf{.857}$	& $.845$  & $.292$& $.525$	& $.937$\\ \hline
BiRank$_r$ & $.406$& $.089$   & $.947$ &$.364$ & $.381$	& $.958$ & $.062$& $.271$	& $.755$ & $.205$& $.714$	& $.770$ & $.170$& $.305$	& $.910$\\
RB-BiRank$_r$  & $.500 $ & $.110$   & $.941$ &$\textbf{.500}$& $\textbf{.524}$	& $.963$ & $\textbf{.113}$& $\textbf{.492}$	& $.827$ & $\textbf{.247}$& $\textbf{.857}$ & $\textbf{.951}$ & $\textbf{.311}$& $\textbf{.559}$ & $\textbf{.949}$ \\ \hline
BiRank$_t$    &$.420$ & $.092$ & $\textbf{.952}$ &$.318$  & $.333$	& $.956$ & $.047$ & $.203$ & $.744$ &$.192$& $.667$	& $.762$ & $.123$& $.220$	& $.901$\\
RB-BiRank$_t$  &$\textbf{.522}$& $\textbf{.115}$ & $.948$ & $.409$ & $.426$	& $.961$ & $.105$& $.458$ & $.824$ &$.205$& $.714$	& $.944$ & $.283$& $.508$	& $.947$ \\ \hline
QRep &$.326$& $.072$   & $.943$ &$.272$& $.286$	& $.975$ & $.043$& $.186$ & $.736$ &$.164$& $.571$	& $.828$ & $.113$ & $.203$	& $.898$\\
RB-QRep &$.514$& $.113$   & $.943$ &$.409$ & $.426$	& $.920$ & $.097$& $.424$ & $\textbf{.828}$ & $.205$& $.714$	& $.856$ & $.274$& $.492$	& $.943$ \\ \hline
BRGM &$.007$& $.002$   & $.678$ &$.091$& $.095$   & $.770$ & $-$ & $-$  & $.581$ & $.014$& $.048$	& $.650$ & $-$	 &   $-$ & $.703$\\
RB-BRGM &$.014$& $.003$ & $.537$ &$.091$ & $.095$ & $.591$ & $-$ & $-$  & $.501$ &$.095$ & $.651$ & $.027$ &  $-$	& $-$ & $.534$ \\

\bottomrule
\end{tabular}
}
\label{tab:performance-1}
\end{table*}
The deviation to the ideal ranking is then defined as
\begin{equation}
\sigma(L)=\sqrt{\frac{1}{S}\sum_{i=1}^{S}(n_{i}-n^{0})^{2}},
\label{(11-2)}
\end{equation}
with $n_i$ the number of items reaching at the top $L$ rank in group $i$. The imbalance is then given as the ratio of the actual standard deviation over the ideal standard deviation:
\begin{equation}
Imbalance=\left|\frac{\sigma}{\sigma_0}-1\right|.
\end{equation}
\label{(11-3)}
A value close to 0 that corresponds to a time balanced ranking, and the more it differs from 0, the more the ranking is time biased.

Additionally to the imbalance metric, we use the following standard metrics.
\textbf{Precision} and \textbf{Recall} measure the algorithm's accuracy, focusing on
the top of the ranking list.
\textbf{NDCG},
the Normalized Discounted Cumulative Gain~\cite{ndcg} is a popular ranking quality measure. Compared to
Precision and Recall, NDCG account for the position of the item in the ranking list.
\textbf{AUC}~\cite{davis2006relationship} is a common statistical significance metric of information retrieval method, and is used
to evaluate the whole list.

\subsection{Baselines}
\label{sec:baseline}
In order to prove our balance framework's performance and universality, we selected three other important bipartite ranking algorithms.
 These algorithms are derived from the HITS algorithm \cite{kleinberg1999authoritative}. Note that most of these methods propose initial values for the fitness and reputation value based on a query or additional information. We restrict the methods to the case where there is no additional information.
In this work, we use three more methods as our baseline, they are \textbf{BiHITS}~\cite{deng2009generalized}, \textbf{QRep}~\cite{L2016Vital} and \textbf{BGRM}~\cite{rui2007bipartite}.

\section{Results and Discussion}
\label{sec:results}
In this section, we compare the results of the algorithms and evaluate their ranking performance with the four metrics described previously. The code and data used in this work are available online\footnote{https://github.com/Joey5555/Rebalance-project}.

\subsubsection{Evaluation of important node recognition}
\begin{figure}[!htbp]
\includegraphics[width=\linewidth,height=2in]{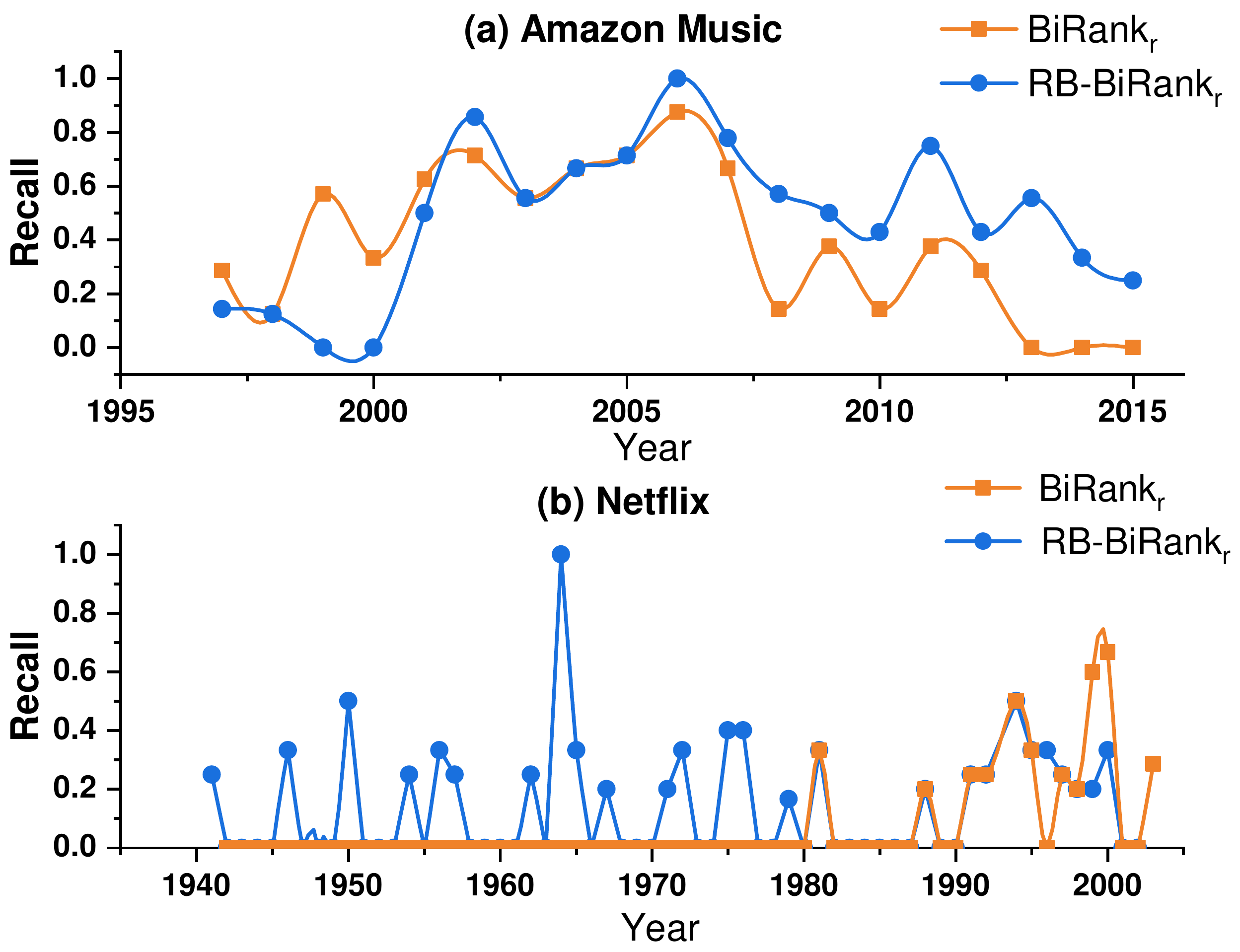}
\caption{Recall for rating BiRank before and after applying the rebalance method for (a) Amazon Music (Grammy) and (b) Netflix (IMDb) datasets.
}
\label{fig:qrep_recallbyyear}
\end{figure}

Table~\ref{tab:performance-1} shows the results obtained by each algorithm under three evaluation metrics, Recall, Precision and AUC for the ground truth built from awards. It can be seen that for all datasets, all the Recall and Precision scores are increased of the rebalance algorithms, better than its corresponding original algorithm, while the value of AUC is maintained at a relatively stable level. The only exception with BRGM algorithm, for which the Recall score is anyway small. Among them, the RB-BiRank$_t$ algorithm has the best Recall value and Precision value for Amazon Music. The RB-BiRank$_r$ algorithm performs best on Precision and Recall for MovieLens and Netflix.
\begin{figure}[!htbp]
\centering
\includegraphics[width=\linewidth]{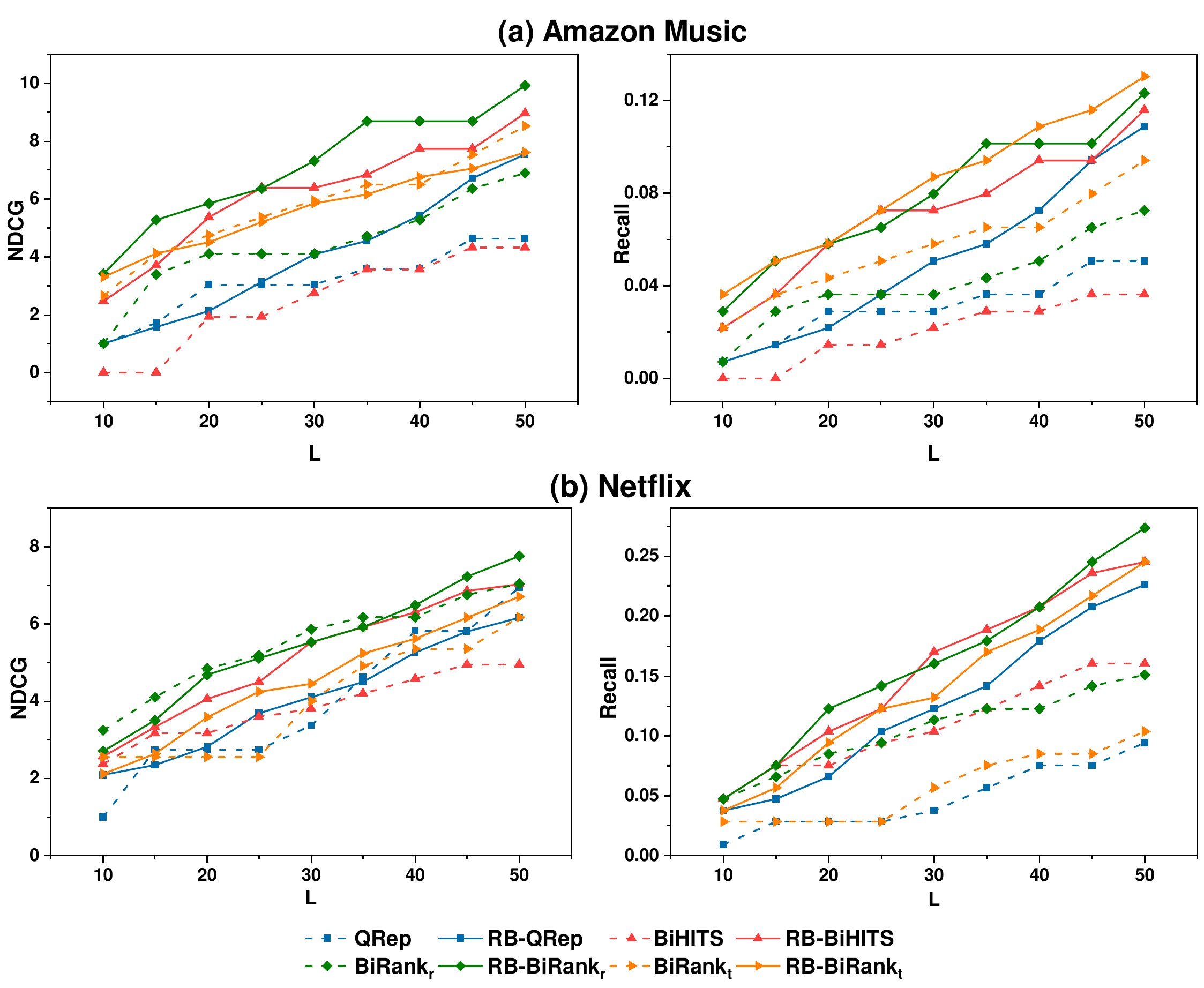}
\caption{NDGC and Recall values as a function of the length of the results list $L$ for (a) Amazon Music (Grammy) and (b) Netflix (IMDb) datasets. The dashed line represents the performance of the algorithms before the rebalance process, and the solid lines after the rebalance process.}
\label{fig:ndcg_recall_new}
\end{figure}
Figure~\ref{fig:qrep_recallbyyear} shows the value of Recall for Birank$_r$ and its rebalance counterpart, RB-Birank$_r$ for Amazon Music and Netflix.
In order to compute the Recall for each year, we use the data up to the limit date of the Awards. For instance, to be nominated for the Oscar of year $T$, a movie needs to open at the latest on the 31st of December of year $T-1$. Then, the ground truth list is composed of the movies awarded at year $T$. For Amazon Music, the benefit is clear for the recent years, while a bit outperformed by the original method in the early years. For Netflix on the other hand, the improvement of the Recall metric compared to the original method is in the early years. This shows that this rebalance process effectively rebalance the score across the years, not favoring early or later years.

\begin{figure}[!htbp]
    \centering
    \includegraphics[width=\linewidth]{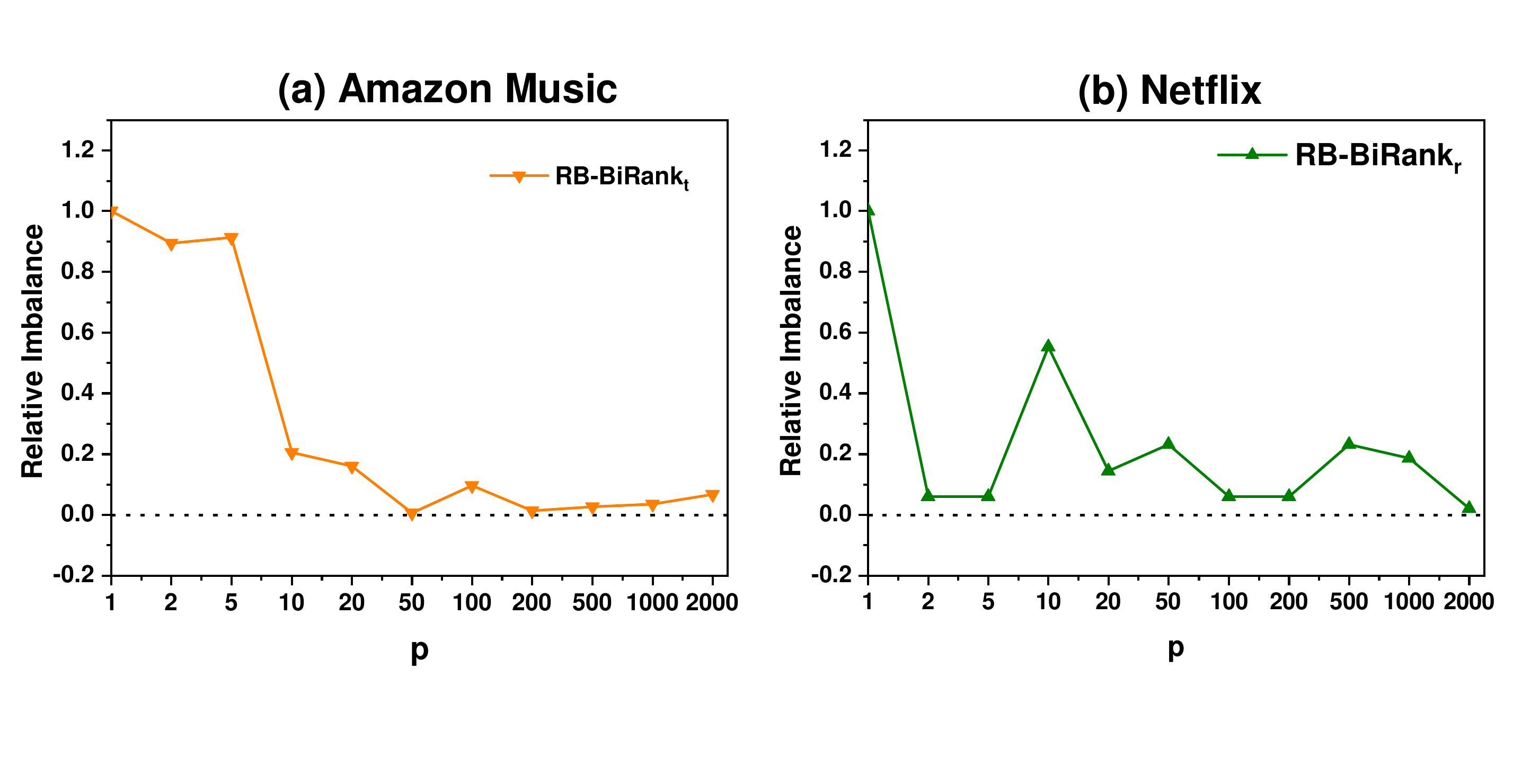}
    \caption{Influence of the value of the number of groups $p$ on the imbalance value of the top-1\% items on (a) Amazon Music (Grammy) and (b) Netflix (IMDb) datasets. The Relative imbalance is the value of the imbalance after the rebalance process over the original imbalance value.}
    \label{fig:delta_p_balance}
\end{figure}
In Figure~\ref{fig:qrep_recallbyyear}(a), it can be seen that after 2001, all the Recall value of RB-Birank$_r$ are better than Birank$_r$'s, and after 2010, each year's recall value of Birank$_r$ is very small, even zero, while the RB-Birank$_r$ keeps a relatively stable level. And in Figure~\ref{fig:qrep_recallbyyear}(b), it is interesting that RB-Birank$_r$ performs both better for older items than its no-rebalance counterpart while being outperformed for newer items. This shows that this modification not only deals for the bias towards the older item, but can also benefit to recent ones.
The trends are similar to other algorithms, hence we only show Birank$_r$ and RB-Birank$_r$.

We also evaluate the impact of the length of the results list $L$ on Recall and NDGC. The results are shown on Figure~\ref{fig:ndcg_recall_new}. For the Amazon Music dataset, we see that the rebalance method improves the performance of most algorithms, independently of the length of the results list. The only exception to this is the algorithm BiRank$_t$ evaluated with the NDCG metric. Similar observations can be made for the Netflix dataset for BiRank$_r$. These results show that the rebalance method improves the results for the top of the ranking list, but also consistently when going further into the list.

\subsubsection{Evaluation of time balance}

In order to compare the time balance of the rankings for the 10 algorithms, we first divide all the items into $40$ groups according to their time. The first group is composed of the oldest items, and the 40th group is the newest items. Then, we take the list of 1\% top rated items for each algorithm, and we count how many of these items belong to each group. Table~\ref{tab:rec_balance} shows the imbalance value of all algorithms. The comparison of imbalance before and after the rebalance process is very clear. The rebalance algorithm can significantly reduce the time imbalance of every algorithm considered in this work. The RB-BiHITS algorithm has the best performance, as its imbalance value is the closest to zero on both Amazon Music and MovieLens data sets. This means that in the result of RB-BiHITS, each groups' number distribution of top item is the closest to hypergeometric distribution.
\begin{table}[!htpb]
\small
\centering
\caption{The imbalance value of the rankings obtained with algorithms.}
\label{tab:rec_balance}
\vspace{-8pt}
\begin{tabular}{c|c|c|c}
\toprule
\multirow{1}{*}{\textbf{Method}} & \multicolumn{1}{c|}{\textbf{Amazon Music}} & \multicolumn{1}{c|}{\textbf{MovieLens}} & \multicolumn{1}{c}{\textbf{Netflix}} \\
\midrule
 BiHITS	& $7.06$ & $0.02$ & $0.30$ \\
 RB-BiHITS & $\textbf{0.14}$ & $0.04$ & $\textbf{0.03}$\\ \hline
 BiRank$_r$	   & $5.57$ & $\textbf{0.01}$ & $0.44$ \\
RB-BiRank$_r$  & $0.10$ & $0.11$ & $\textbf{0.03}$\\ \hline
BiRank$_t$ & $5.10$ & $0.04$ & $0.60$\\
RB-BiRank$_t$  & $0.18$ & $0.06$ & $\textbf{0.03}$\\ \hline
 QRep         & $5.24$ & $0.26$  & $0.48$\\
RB-QRep      & $0.15$ & $0.10$  & $0.18$\\ \hline
 BRGM         & $2.21$ & $\textbf{0.01}$ & $0.09$\\
RB-BRGM     & $1.45$ & $0.06$  & $0.30$\\ \hline
\end{tabular}
\vspace{-10pt}
\end{table}
\subsubsection{The influence of $p$ on imbalance}

In order to study the influence of the number of group $p$ on imbalance value, we compute the value of the imbalance for different values of $p$, ranging from 1 to 2000. We show the results in Figure~\ref{fig:delta_p_balance} for the rebalanced BiRank algorithms. The $y$-axis on the figure represents the relative imbalance: the $imbalance$ before over the $imbalance$ after applying the rebalance method. It can be seen that on Figure~\ref{fig:delta_p_balance}(a) that for Amazon Music, the imbalance stabilizes and remain small after $p=10$. For Netflix in Figure~\ref{fig:delta_p_balance}(b), the algorithm is able to keep a low relative imbalance after 20. This result is important as it shows that the value of imbalance stays quite small at $p$, and so does not need to be finely tuned.
\begin{table}[!htbp]
\small
\centering
\setlength{\tabcolsep}{1.0mm}
\caption{Top-10 albums in the ranking lists of RB-BiRank$_r$ and BiRank algorithms.}
\makebox[\linewidth][c]{
\begin{tabular}{l l | l l}
\toprule
\multicolumn{2}{c}{\textbf{\underline{RB-BiRank$_r$}}} & \multicolumn{2}{c}{\textbf{\underline{BiRank$_r$}}}  \\
\rule{0pt}{5pt}\textbf{Album Name}  & \multicolumn{1}{r|}{\textbf{Musician}} &  \textbf{Album Name} & \multicolumn{1}{r}{\textbf{Musician}} \\
\midrule
      Death Magnetic & Metallica(2008) & \multicolumn{1}{l}{\begin{tabular}[l]{@{}l@{}}The Dark Side \\ of the Moon\end{tabular}}  & Pink Floyd(1973) \\ \hline
      \rule{0pt}{10pt} 1 & The Beatles(2000) &  The Beatles & The Beatles(1968) \\ \hline
      \rule{0pt}{10pt} Lateralus & Tool(2001) & Abbey Road & The Beatles(1969) \\ \hline
      Fallen & Evanescence(2003) & \multicolumn{1}{l}{\begin{tabular}[l]{@{}l@{}}Sgt. Pepper's\\ Lonely Hearts\\ Club Band\end{tabular}} & The Beatles(1967) \\ \hline
      \rule{0pt}{10pt} Kid A & Radiohead(2000)  & 1 & The Beatles(2000) \\ \hline
     \multicolumn{1}{l}{\begin{tabular}[l]{@{}l@{}}Confessions on\\ a Dance Floor\end{tabular}}  & Madonna(2005) &  \multicolumn{1}{l}{\begin{tabular}[l]{@{}l@{}}Master of \\ Puppets\end{tabular}}  & Metallica(1986) \\ \hline
     \multicolumn{1}{l}{\begin{tabular}[l]{@{}l@{}}Come Away \\ with Me\end{tabular}}   & Norah Jones(2002) & OK Computer & Radiohead(1997) \\ \hline
    \multicolumn{1}{l}{\begin{tabular}[l]{@{}l@{}}The\\ Emancipation\\  of Mimi\end{tabular}} & Mariah Carey(2005) & Nevermind & Nirvana(1991) \\ \hline
    \rule{0pt}{10pt} St. Anger & Metallica(2003) & Revolver & The Beatles(1966)\\ \hline
     Music & Madonna(2000) &  \multicolumn{1}{l}{\begin{tabular}[l]{@{}l@{}}Wish You \\ Were Here\end{tabular}} & Pink Floyd(1975) \\
  \bottomrule
\end{tabular}
}
 \label{tab:case_study}
\end{table}
\subsubsection{Case study}
In Table~\ref{tab:case_study}, we illustrate the effect of the rebalance process on the top of the list. To this end, we run the algorithm on the amazon dataset, and we list the top-10 albums ranked by the original and the rebalanced version of BiRank$_r$. The effect of the rebalance is clear: there are more recent albums than with the original algorithm. Moreover, the diversity is higher with the rebalanced version, as the original algorithm has only five different musicians in the top-10 list, while the rebalanced list has eight different musicians.

\section{Conclusion and Perspective}
\label{sec:conclusion}
Due to the huge amount of content available online, it is difficult for an individual to find high quality content. The online companies might favor the popular contents for advertisement purpose, despite the quality of the content. Then, people may spend a tremendous amount of time searching for the contents of good quality. This is where ranking algorithms such as the one developed in this work can prove useful.

In this work, we proposed a novel and general balance framework to balance the ranking results without losing the accuracy but obtain rankings that include items from all time periods. We showed that this method can be easily applied to all ranking algorithms, and that it constantly improved the accuracy of the algorithms. Moreover, it does not only push relevant items at the top of the rankings, but we showed that the quality of the rankings is also improved if we look further down the list. We could not single out one algorithm which performed constantly better than the others, some algorithms have better performance on specific datasets.

Our work combined a general method and the evaluation of the outcomes with quantitative metrics based on ground truth established by the experts of the domain. In the future study, the balance framework could be applied to the more general domain such as economic systems and those systems without ground truth. The proposed method also could be a good tool to evaluate the time bias of new ranking algorithms.

\clearpage
\newpage
\bibliographystyle{aaai}
\bibliography{aaai2020}
\end{document}